# Nonlinear wavefront control by geometric-phase dielectric metasurfaces: Influence of mode field and rotational symmetry

*Bingyi Liu[#1,2], Basudeb Sain[#2], Bernhard Reineke[#2], Ruizhe Zhao[3], Cedrik Meier[2], Lingling Huang[*3], Yongyuan Jiang[*1,4], and Thomas Zentgraf[*2]*

[1]School of Physics, Harbin Institute of Technology, Harbin 150001, China

[2]Department of Physics, Paderborn University, Warburger Straße 100, D-33098 Paderborn, Germany

[3]School of Optics and Photonics, Beijing Institute of Technology, Beijing 100081, China

[4]Key Lab of Micro-Optics and Photonic Technology of Heilongjiang Province, Harbin 150001, China
Key Laboratory of Micro-Nano Optoelectronic Information System of Ministry of Industry and Information Technology, Harbin 150001, China
Collaborative Innovation Center of Extreme Optics, Shanxi University, Taiyuan 030006, China

[#]These authors contributed equally to this work.

Keywords: nonlinear optics, metasurface, geometric-phase, hologram

Nonlinear Pancharatnam-Berry phase metasurfaces facilitate the nontrivial phase modulation for frequency conversion processes by leveraging photon-spin dependent nonlinear geometric-phases. However, plasmonic metasurfaces show some severe limitation for nonlinear frequency conversion due to the intrinsic high ohmic loss and low damage threshold of plasmonic nanostructures. Here, we systematically study the nonlinear geometric-phases associated with the third-harmonic generation process occurring in all-dielectric metasurfaces, which are composed of silicon nanofins with different in-plane rotational symmetries. We find that the wave coupling among different field components of the resonant fundamental field gives rise to the appearance of different nonlinear geometric-phases of the generated third-harmonic signals. The experimental observations of the nonlinear beam steering and nonlinear holography realized in this work by all-dielectric geometric-phase metasurfaces are well explained with our developed theory. Our work offers a new physical picture to understand the nonlinear optical process occurring at nanoscale dielectric resonators and will help in the design of nonlinear metasurfaces with tailored phase properties.

**1. Introduction**

Optical metasurfaces offer the ultimate solution for modern diffractive optical elements design by leveraging the artificially tailored nanostructures of subwavelength geometries, which facilitate the flexible control of the phase, amplitude, and polarization of the light with an ultra-compact profile.[1,2] Since the propagation of the light is substantially determined by the local phase distribution of the wavefront, metasurface structures with full-phase transmitted or reflected phase discontinuities are highly desired for realizing numerous optical functionalities.[3-8] In particular, Pancharatnam-Berry phase or geometric-phase metasurfaces simplify the optical wavefront engineering by utilizing the linear Pancharatnam-Berry phase, which can be precisely controlled by locally rotating the nanostructures.[9-12]

At the very beginning, plasmonic metasurfaces gained its popularity for the simplicity of obtaining an effective electric and magnetic response with spatially tailored induced surface currents.[13] Besides that, due to the strong local field enhancement near the surface of the plasmonic nanostructures and free of restrictions by the phase-matching condition, plasmonic metasurfaces with boosted nonlinear generation efficiency show great applicability in nonlinear nanophotonics.[14,15] However, the plasmonic materials inevitably introduce large ohmic losses and suffer from low damage thresholds while the all-dielectric materials offer a promising alternative. Benefiting from the multipole Mie resonances supported by the dielectric nanostructures, all-dielectric metasurfaces with greatly decreased dissipation loss and robustness against high power lasers are widely applied to the efficient linear and nonlinear optical wavefront manipulation.[16-21] In addition, the sharp Fano resonances supported by the all-dielectric metasurfaces[22-25] and the excitation of the nonradiative modes in the structures[26-28] give rise to the extreme local field confinement and provide stronger light-matter interaction for the efficient nonlinear signal generations.

Recently, the nonlinear optical phase obtained with designed nanostructures has attracted tremendous research interest for the convenience they provide in controlling the nonlinear optical waves. Besides the nonlinear Huygens metasurfaces,[21,29,30] the photon-spin dependent nonlinear geometric-phase opens another avenue to the efficient phase modulation of the photons generated during the frequency conversion process. The pioneering work that takes advantage of the hybrid metasurfaces based on plasmonic structures and multiple-quantum-wells in semiconductors reveals the possibility of exploiting the spatially varied nonlinear geometric-phase for advanced nonlinear wavefront control with high-efficiency.[31-33] In addition, by revisiting the selection rule that holds for the nonlinear crystals, it was revealed that plasmonic nanostructures with specific structural symmetry also show similar selection behavior for the harmonic generation processes.[34-38] In nonlinear geometric-phase metasurfaces, recently reported works mostly focusses on implementing plasmonic nanostructures while studies on all-dielectric nonlinear geometric-phase metasurfaces are still missing. Here, we theoretically and experimentally study the nonlinear geometric-phases associated with the third-harmonic generation (THG) process occurring in all-dielectric metasurfaces, which are composed of silicon nanofins with different in-plane rotational symmetries. In our previous work,[39] we experimentally exploited the potential of the nonlinear geometric-phase for spin-multiplexed nonlinear holograms using silicon nanofin of two-fold symmetry. However, the underlying physics of the geometric-phase in a nonlinear process using such spatially extended nanofins was only based on some phenomenological assumptions. Here, we analytically develop a complete theory to reveal the underlying physics for the generation of nonlinear geometric-phase and describe the influence of the structural symmetry and local optical mode structure on the possible phase factors. Based on the full-wave simulation and experimental measurements, we test our theoretical analysis by realizing nonlinear beam steering and nonlinear holography based on such metasurfaces.

## 2. Results and discussion
### 2.1. Nonlinear polarization excited in the dielectric geometric-phase metasurfaces

For plasmonic geometric-phase metasurfaces, the harmonic generation process can be phenomenally understood as the result of the excitation of the effective nonlinear surface currents which substantially carry the spin-dependent nonlinear geometric-phases.[32] Moreover, the nonlinear response of the plasmonic metasurfaces is usually modeled as an equivalent surface with effective tensor nonlinear susceptibilities where the wave coupling and phase matching along the propagating direction is generally neglected due to the short propagation distances.[32,33] However, the nonlinear optical signals generated from dielectric nanostructures are built up by the overall radiation of the induced volume nonlinear polarization. Due to the large aspect ratio and the high refractive index of the dielectric structures, the propagation of the light confined inside the structures makes the wave coupling effect conspicuous. Hence, the nonlinear optical process occurring in dielectric nanostructures will inevitably involve the wave coupling among different field components of the resonant fundamental wave (FW), and the radiated nonlinear waves can carry additional geometric-phases beyond what has been predicted in their plasmonic counterpart.

A conceptual image of the used metasurface is shown in **Figure 1**, which illustrates the effect of anomalous third-harmonic (TH) beam steering exhibited by the nonlinear metasurface with designed spatially linear phase modulation. When a circularly polarized FW normally impinges onto such a metasurface, the generated LCP and RCP TH signals, which are modulated with the spin-dependent nonlinear geometric-phases, will be deflected into different spatial directions. In our work, we always assume planar circularly polarized FW propagating along the normal of the metasurface, which is collinear with the z-axis of the laboratory frame, illuminates the metasurface from the substrate, see Figure 1. For the THG processes occurring in the dielectric metasurface, the wave coupling between different components of the confined FW field can be described through the effective third-order susceptibility tensor $\chi_{ijkl}^{(3)\text{eff}}$, where *i*, *j*, *k*, and *l* refer to the cartesian coordinates x, y, and z, respectively. Because the nonlinear geometric-phase is carried by the circularly polarized TH light waves, it is more convenient to study the nonlinear response in the circular polarization basis. The effective third-order susceptibility tensor in the circular basis can be obtained by a coordinate transformation, which finally gives $\chi^{(3)\text{eff}}_{\alpha\beta\gamma\delta} = \sum_{ijkl} \chi^{(3)\text{eff}}_{ijkl} \Lambda^{\alpha}_i \Lambda^{j}_{\beta} \Lambda^{k}_{\gamma} \Lambda^{l}_{\delta}$. Here, $\alpha$, $\beta$, $\gamma$, and $\delta$ refer to the circular coordinates $L$, $R$, and z, and $\Lambda^{\alpha}_i$ is the element of the transformation matrix between the circular basis and the cartesian basis. When rotating the in-plane orientation of the dielectric nanostructure, the rotated nonlinear polarization will differ from the unrotated nonlinear polarization in the laboratory frame by an additional phase, which gives rise to the nonlinear geometric-phase of the radiated TH signals. As shown in Figure 2(a), when the silicon nanofin is counterclockwise rotated by an angle $\theta$, the effective third-order susceptibility tensor of the nanostructure in the circular basis can be written as $\chi^{(3)\text{eff}}_{\alpha'\beta'\gamma'\delta'} = \chi^{(3)\text{eff}}_{\alpha\beta\gamma\delta} R^{\alpha'}_{\alpha} R^{\beta}_{\beta'} R^{\gamma}_{\gamma'} R^{\delta}_{\delta'}$, here the prime refers to the local frame, and $R^{\alpha'}_{\alpha}(\theta)$ is the element of the rotation matrix between the local and laboratory frame in the circular basis. Therefore, the nonlinear optical process that involves the wave coupling of specific resonant FW field components correspondingly carries the geometric-phase contributed by the local and laboratory frame rotation transformation, i.e., $R^{\alpha'}_{\alpha} R^{\beta}_{\beta'} R^{\gamma}_{\gamma'} R^{\delta}_{\delta'}$. Considering only the left-handed (LCP) and right-handed (LCP) TH signals that arise from the forward THG processes, we analytically obtain the corresponding volume third-order nonlinear polarizations as:

$$P^{(3)\text{eff}}_L = \varepsilon_0 \left[ \chi^{(3)\text{eff}}_{LLLL} e^{2i\theta} \left(E^L\right)^3 + \chi^{(3)\text{eff}}_{LRRR} e^{-4i\theta} \left(E^R\right)^3 \right] \\ + 3\varepsilon_0 \left[ \chi^{(3)\text{eff}}_{LLLR} E^R \left(E^L\right)^2 + \chi^{(3)\text{eff}}_{LLRR} e^{-2i\theta} E^L \left(E^R\right)^2 \right],$$ (1a)

$$P_R^{(3)\text{eff}} = \varepsilon_0 \left[ \chi_{RRRR}^{(3)\text{eff}} e^{-2i\theta} \left(E^R\right)^3 + \chi_{RLLL}^{(3)\text{eff}} e^{4i\theta} \left(E^L\right)^3 \right]$$
$$+ 3\varepsilon_0 \left[ \chi_{RRRL}^{(3)\text{eff}} E^L \left(E^R\right)^2 + \chi_{RRLL}^{(3)\text{eff}} e^{2i\theta} E^R \left(E^L\right)^2 \right].$$
(1b)

where $P_L^{(3)\text{eff}}(\mathbf{r},\theta,3\omega)$ and $P_R^{(3)\text{eff}}(\mathbf{r},\theta,3\omega)$ are the LCP and RCP components of the effective volume nonlinear polarizations, $\mathbf{r}$ is the location of an arbitrary spatial point in the dielectric metasurface, $\theta$ is the rotation angle of the silicon nanofin and $3\omega$ is the TH angular frequency. $E^R(\mathbf{r},\theta,\omega)$ and $E^L(\mathbf{r},\theta,\omega)$ refer to the RCP and LCP component of the effective FW field confined in the dielectric metasurface, respectively. A detailed derivation of Equation 1 can be found in Supporting Information.

According to Equation 1, for the THG process occurring in the silicon metasurfaces, it is theoretically possible to allow the existence of nonlinear polarizations carrying the geometric-phase of $\exp(\pm 2i\theta)$ and $\exp(\pm 4i\theta)$. Therefore, the far-field TH waves generated by these nonlinear polarizations would carry the corresponding geometric-phase as well. The intensity of the TH waves contributed by each nonlinear polarization term is determined by the value of the effective nonlinear susceptibilities and field intensity of the confined FW components. In addition, Equation 1 provides some insight into the harmonic generation process occurring in the nanofins. For example, considering the THG process occurring in the geometric-phase metasurface composed by silicon nanofins of two-fold (C2) in-plane rotational symmetry, whose lattice constant is small enough compared to the FW wavelength. In this case, the metasurface can be equivalently treated as a thin effective nonlinear biaxial crystalline film, which is sandwiched between the substrate and air. With such an approach, the FW field confined inside the dielectric metasurface sample can be equivalently treated as the fundamental waveguide mode supported by a thin film. The fundamental waveguide mode can be decomposed as the superposition of forward and backward traveling plane waves propagating along the z-axis. In this way, the nonlinear polarization term $\varepsilon_0 \chi_{LLLL}^{(3)\text{eff}} e^{2i\theta} \left(E^L\right)^3$, which contributes to the LCP TH wave carrying the geometric-phase of $\exp(2i\theta)$, can be understood as the wave coupling among the plane LCP total FW field components propagating in the forward direction. Similarly, the backward propagating total FW fields will also generate the TH signals in the backward direction following similar processes illustrated in Equation 1 (Supporting Information).

It is well known that the selection rule verified by the plasmonic nanoantenna with *m*-fold in-plane rotational symmetry gives the allowed harmonic generation orders by $n = lm \pm 1$ when the metasurface sample is pumped by a circularly polarized FW.[34] Here, *l* is an arbitrary integer and '+' and '−' signs refer to the co-polarized and cross-polarized harmonic signals, respectively. Therefore, in this work, we primarily focus on the THG processes occurring in the silicon nanofins with one-fold (C1), two-fold (C2) and four-fold (C4) in-plane rotational symmetry and how the structural symmetry influences the spin and corresponding geometric-phases carried by the TH signals.

## 2.2. Nonlinear response of a geometric-phase metasurface composed by silicon nanofins with C2 in-plane rotational symmetry

To simplify our analysis, we first study the properties of the TH signals generated from silicon nanofins with C2 in-plane rotational symmetry. It is well known that Neumann's principle determines the dipole allowed nonlinear optical processes for structures with specific symmetry. Therefore, the effective volume nonlinear polarizations excited inside the silicon nanofins with C1 and C2 in-plane rotational symmetries are analytically given by Equation 1. Next, we conduct the full-wave simulations to facilitate the numerical study of the nonlinear response of the silicon nanofin array. The nanofin is made of amorphous silicon (α-Si) and placed on the low-refractive-index silicon dioxide ($SiO_2$) substrate, and the surrounding medium of the above half-space is assumed to be air. In our work, the silicon nanofin is designed to be of a subwavelength scale to ease the wavefront phase encoding of the TH signals. As shown in Figure 2(a), the height, width, and length of the silicon nanofin are selected as $h = 620$ nm, $w = 150$ nm and $l = 300$ nm, respectively. The period of the nanofins is selected as $p = 380$ nm to suppress the unwanted higher-order diffractions of the radiated TH signals and thereby increases the uniformity of the reconstructed wavefront. In all our simulations, we assume the metasurface is normally illuminated by an LCP FW. The nonlinear response of the silicon nanofin is modeled through the nonlinear polarization excited inside the structures. For a silicon nanofin array of which all nanofins have a rotation angle of 0°, Figure 2(b) shows the LCP and RCP components of the confined FW field excited at 1300 nm as well as the corresponding TH fields. Due to the subwavelength geometry of the silicon nanofin, there is no resonant excitation with strong field confinement inside the nanofin at the infrared FW, therefore the FW field stimulated in the nanofin can be generally treated as the fundamental waveguide mode composed of the forward and backward propagating plane waves, which offers a straightforward picture for understanding the wave coupling process in the all-dielectric metasurface. In addition, according to the simulation, the incident FW is highly trapped in the low-index air gap among the silicon nanofins, which indicates the strong coupling among the neighboring nanofins. Figure 2(c) shows the LCP, RCP and the overall forward THG conversion efficiency of the silicon nanofin array for different rotation angles of the nanofins. Here, the THG conversion efficiency is defined by the ratio between the power of the forward TH signal and the input FW power. It can be seen from the figure that the THG conversion efficiency oscillates periodically when locally rotating nanofins. Such a result can be well understood through the variation of the coupling strength among neighboring units, which is equivalent to alter the effective refractive index of the fast and the slow axis of the metasurface sample. Therefore, the total FW field excited in metasurface would correspondingly change with different local orientation and finally leads to the intensity fluctuation of the output TH signals. Figure 2(d) further demonstrates the LCP, RCP and the overall forward THG conversion efficiency spectrum of the silicon nanofin array ($\theta = 0°$) for FW wavelengths between 1150 nm and 1500 nm. Note that the overall THG conversion efficiency increases at smaller FW wavelengths due to stronger localized FW field intensity inside the nanofins, and the intensity of LCP and RCP TH signals will vary with the FW wavelength as well. Especially when the FW wavelength is larger than 1450 nm, the intensity of cross-polarized TH waves is almost zero. It should be mentioned that the low conversion efficiency of co- or cross-polarized TH signals would correspond to the large fluctuation of the geometric-phase carried by them, the mechanism can be understood with Equation 1. Take the LCP FW illumination as an example, generally, the nonlinear polarization $\varepsilon_0 \chi_{LLLL}^{(3)\text{eff}} e^{2i\theta}$ and $\varepsilon_0 \chi_{RLLL}^{(3)\text{eff}} e^{4i\theta}$ are the main terms while the influence of other nonlinear polarizations on the overall TH outputs is negligible. Therefore, the low conversion efficiency of LCP or RCP TH signals suggests the low intensity of above two nonlinear polarizations as well. In this case, the contribution of other nonlinear polarization terms would be obvious and thereby distorts the geometric-phases carrying by main terms (Supporting Information).

Based on the silicon nanofin shown in Figure 2(a), we further simulate the anomalous refractions of the TH signals generated from the metasurface with a gradient nonlinear geometric-phase change along the x-direction. By gradually rotating the silicon nanofin along the x-direction, a spin-dependent surface phase gradient is obtained that diffracts the TH beams at different angles. In our simulation, the FW wavelength is selected as 1300 nm, and the gradient-phase metasurface period is composed of 36 units. Figure 2(e) and (f) show the simulated geometric-phase shifts of the LCP and RCP TH signals for the 36 nanofins along the x-direction. Based on the nonlinear polarization given by Equation 1, the TH waves radiated from the locally rotated silicon nanofins of C2 in-plane rotational symmetry simultaneously carry the geometric-phase of $\exp(\pm 2i\theta)$ and $\exp(\pm 4i\theta)$. According to our simulation result, the geometric-phase associated with the LCP and RCP TH signals is approximately proportional to $\exp(2i\theta)$ and $\exp(4i\theta)$, respectively, which indicates that, in this scenario, $\varepsilon_0 \chi_{LLLL}^{(3)\text{eff}} e^{2i\theta}$ and $\varepsilon_0 \chi_{RLLL}^{(3)\text{eff}} e^{4i\theta}$ are the two main THG processes supported by the dielectric metasurface. Figure 2(g) and (h) show the obtained field distributions of the transmitted LCP and RCP TH signals in the x-z-plane. From the field plots, we obtain the refraction angles of 1.78° and 3.67°, which are close to the design values of 1.82° and 3.63°, respectively.

**2.3. Nonlinear response of the geometric-phase metasurface composed of silicon nanofins with C1 and C4 in-plane rotational symmetry**

For further demonstration of the selection rule, we study the geometric-phases associated with the TH waves generated from the silicon nanofins with C1 and C4 in-plane rotational symmetry. Figure 3(a) illustrates the used C1 silicon nanofins. In our design, the C1 silicon nanofin has a U-shape cross-section, and its geometric parameters are selected as $w$ = 150 nm, $l$ = 300 nm, $d_1$ = 80 nm and $d_2$ = 60 nm. The height and lattice constant of such silicon nanofin are optimized as $h$ = 550 nm and $p$ = 400 nm to maintain higher THG conversion efficiency. By locally rotating the silicon nanofin from 0° to 360° with an angular rotation step of 5°, the THG conversion efficiency will change due to the variation of the coupling among the neighboring units, see Figure 3 (b). The larger fluctuation of THG conversion efficiency indicates stronger coupling among silicon nanofins occurs in our C1 design, and bigger distortion on the nonlinear geometric-phases is expected. In our simulation, the FW wavelength is selected as 1240 nm to ensure that the TH wavelength is bigger than the lattice constant, which suppresses the higher-order diffraction and increases the uniformity of the TH far-field. Figure 3 (c) and (d) are the calculated FW and TH field distribution. Figure 3(e) and (f) illustrate the geometric-phase carried by the co-polarized and cross-polarized TH waves, which show a nearly linear relation of two times and four times of the rotation angle, respectively. This agrees with the nonlinear polarizations given by Equation 1, where $\varepsilon_0 \chi_{LLLL}^{(3)\text{eff}} e^{2i\theta}$ and $\varepsilon_0 \chi_{RLLL}^{(3)\text{eff}} e^{4i\theta}$ are the main THG processes contributed to the far-field

TH signals.

Based on Neumann's principle, for silicon nanofins of C4 in-plane rotational symmetry, its effective volume nonlinear polarizations are given by:

$$P_L^{(3)\text{eff}}(\mathbf{r},\theta,3\omega) = \varepsilon_0 \left[ \chi_{LRRR}^{(3)\text{eff}} e^{-4i\theta} \left(E^R\right)^3 + 3\chi_{LLLR}^{(3)\text{eff}} E^R \left(E^L\right)^2 \right], \quad (2a)$$

$$P_R^{(3)\text{eff}}(\mathbf{r},\theta,3\omega) = \varepsilon_0 \left[ \chi_{RLLL}^{(3)\text{eff}} e^{4i\theta} \left(E^L\right)^3 + 3\chi_{RRRL}^{(3)\text{eff}} E^L \left(E^R\right)^2 \right]. \quad (2b)$$

Here, the TH waves radiated from the locally rotated array silicon nanofin with C4 in-plane rotational symmetry would carry the geometric-phase of $\exp(\pm 4i\theta)$. Figure 4(a) illustrates the geometry of the silicon nanofin of C4 in-plane rotational symmetry, where the width, length, height and the lattice constant are optimized as $w = 130$ nm, $l = 280$ nm, $h = 500$ nm, and $p = 390$ nm, respectively. By normally illuminating the metasurface sample with an LCP FW at 1350 nm, the RCP and LCP components of the FW excited inside the C4 silicon nanofin are shown in Figure 4(b). It is apparent that the amplitude of the RCP components of the total FW field confined inside the silicon nanofin is very small when compared with the LCP components. Hence the intensity of the nonlinear polarization $\varepsilon_0 \chi_{LRRR}^{(3)\text{eff}} e^{-4i\theta} \left(E^R\right)^3$ would be much smaller than the nonlinear polarization $\varepsilon_0 \chi_{RLLL}^{(3)\text{eff}} e^{4i\theta} \left(E^L\right)^3$, which indicates that only cross-polarized TH waves can be obtained in the transmitted far-field.

Figure 4(c) illustrates the geometric-phase carried by the cross-polarized TH waves when locally rotating the C4 silicon nanofin from 0° to 90° with an angular rotation step of 5°, and the geometric-phase is proportional to four times of rotation angle. Figure 4(d) shows the THG conversion efficiency of the co-polarized and cross-polarized TH signals and it is obvious that the co-polarized TH signals almost do not exist. Based on our theoretical study about the C1, C2, and C4 silicon structures, we reach the following conclusion for the selection behavior shown by the THG occurring in the silicon geometric-phase metasurfaces: assuming the metasurface sample is pumped by the FW with spin σ, for C1 and C2 silicon nanofins, the co-polarized and cross-polarized TH waves mainly carry the geometric-phase of $\exp(2i\sigma\theta)$ and $\exp(4i\sigma\theta)$, respectively. The C4 silicon nanofins only support the cross-polarized TH signals carrying the geometric-phase of $\exp(4i\sigma\theta)$ (see Supporting Information).

In Figure 4(e) and (f), the full-wave simulation shows the field distribution of the cross-polarized and co-polarized TH signals radiated from the C4 silicon nanofins with a gradient local geometric-phase modulation. In our full-wave simulation, the FW wavelength is selected as 1350 nm to increase the linearity between the local rotation angle and the geometric-phase, and the designed gradient geometric-phase metasurface is composed by 20 silicon nanofins within one period. According to our calculation, we obtain an anomalous refraction angle of 3.36°, which is close to the theoretical values of 3.31° calculated by generalized Snell's law. In addition, the TH waves radiated from the C4 structure could also be equivalently understood as the superposition of the TH waves radiated from two overlapped C2 nanofins with a 90° intersection angle. In this case, the co-polarized TH wave obtained in the far-field is the superposition of two co-polarized TH waves differing by 180° phase delay and hence interfere destructively. In contrast, the cross-polarized TH wave is the superposition of two cross-polarized TH waves in phase. Therefore, only cross-polarized TH waves can be observed in the far-field due to the constructive interference that occurs for the cross-polarized TH waves radiated from the nanofins.

## 2.4. Experimental verification of the nonlinear beam steering realized by silicon gradient geometric-phase metasurfaces

We proceed to experimentally validate the nonlinear phase manipulation by utilizing the spin-dependent nonlinear geometric-phases associated with the silicon nanofin metasurface. The dielectric metasurface was fabricated on a silica substrate by standard nanofabrication processes of deposition, patterning, lift-off, and etching. Figure 5(a) shows a scanning electron microscopy (SEM) image of the fabricated silicon nanofin gradient geometric-phase metasurface sample composed of C2 silicon nanofins, where one supercell is composed by 20 gradually rotated silicon nanofins with an angular rotation step of 9°. In our design, the geometry parameters of the nanofin are the same as illustrated in Figure 2(a). The experimental setup is schematically shown in Figure 5(b). The FW beam was generated by an optical parameter oscillator (Coherent, Chameleon compact OPO) and passed through a linear polarizer and a quarter waveplate to obtain the desired circularly polarized FW input. Then the FW was focused onto the sample and the TH signal was collected with an objective (Nikon 40x, NA 0.6). The back focal plane was imaged by two lenses to an sCMOS camera (Andor, Zyla 4.2). The polarization state of the collected TH signal was analyzed with another quarter-wave plate and a linear polarizer. Short-pass filters were used to suppress the FW background. Figure 5(c) shows the measured diffraction patterns of the LCP and RCP TH signals when the gradient-phase metasurface sample is illuminated with 50 mW circularly polarized FW at 1200 nm. Take the LCP FW illumination scenario as an example, the measured diffraction angles for the LCP and RCP TH signals are 3.05°±0.01° and 6.09°±0.01°, respectively, which is close to the design values of 3.01° and 6.01°, correspondingly. Figure 5(d) is the intensity profile of the center cutline of the measured TH diffraction pattern. It can be clearly seen here that for both co-polarized and cross-polarized TH signals, we could always observe the zeroth order, which can be explained by the nonlinear polarizations $3\varepsilon_0 \chi_{LLLR}^{(3)\text{eff}} E^R (E^L)^2$ and $3\varepsilon_0 \chi_{RRRL}^{(3)\text{eff}} E^L (E^R)^2$ given in Equation 1. Moreover, for co-polarized TH signals, we could also see an obvious conjugate diffraction order. Take the LCP FW illumination as an example, the bright +1st diffraction order and a visible −1st diffraction order are believed to be contributed by the nonlinear polarization $\varepsilon_0 \chi_{LLLL}^{(3)\text{eff}} e^{2i\theta} (E^L)^3$ and $3\varepsilon_0 \chi_{LLRR}^{(3)\text{eff}} e^{-2i\theta} E^L (E^R)^2$, respectively. For the influence of nonlinear polarization $\varepsilon_0 \chi_{LRRR}^{(3)\text{eff}} e^{-4i\theta} (E^R)^3$, it is negligible and comparable to the high order diffraction contributed by the Bragg scattering of the gradient metasurface (see the weak periodic fluctuations appear in the intensity profile of co-polarized TH signals).

We also fabricate and test the gradient geometric-phase metasurfaces made of C1 and C4 silicon nanofins as shown previously in Figure 3 and 4. Figure 6(a) shows the experimental measurements of the TH diffraction pattern radiated from the gradient metasurface composed by C1 silicon nanofins. Here, one gradient period contains 15 nanofins. A circularly polarized FW, whose wavelength and laser power are selected as 1240 nm and 50 mW, normally illuminates the metasurface sample from the silica substrate. For the LCP FW illumination scenario, the measured diffraction angles of LCP and RCP TH signals are 7.54°±0.01° and 15.29°±0.01°, respectively, which are close to the theoretical values of 7.92° and 16.00°, correspondingly. Based on the intensity profile of the measured TH diffraction pattern, we could see high diffraction efficiency holds for co-polarized TH waves where nonlinear polarization $\varepsilon_0 \chi_{LLLL}^{(3)\text{eff}} e^{2i\theta} (E^L)^3$ or $\varepsilon_0 \chi_{RRRR}^{(3)\text{eff}} e^{-2i\theta} (E^R)^3$ dominates co-polarized TH far-field. In addition, the cross-polarized TH diffraction pattern obtained in, e.g., LCP FW illumination scenario, shows obvious existence of diffracted TH waves carrying the geometric-phase proportional to two times of rotational angle and visible zeroth-order diffraction. This indicates the considerable contribution of nonlinear polarizations $3\varepsilon_0 \chi_{RRLL}^{(3)\text{eff}} e^{2i\theta} E^R (E^L)^2$ (+1st order) and $3\varepsilon_0 \chi_{RRRL}^{(3)\text{eff}} E^L (E^R)^2$ (0th order) that beyond the expected nonlinear polarization $\varepsilon_0 \chi_{RLLL}^{(3)\text{eff}} e^{4i\theta} (E^L)^3$ (+2nd order).

Figure 6(b) illustrates the measured TH diffraction pattern when the gradient metasurface constructed by C4 silicon nanofins are normally illuminated with a circularly polarized FW whose wavelength and laser power are set as 1340 nm and 50 mW. Here, one gradient period contains 10 units. It is apparent that anomalous diffracted TH spots can only be observed in cross-polarized TH waves, for co-polarized TH signals, we could only obtain the zeroth-order diffraction spot. Considering the LCP FW illumination case, the measured anomalous refracted angle of RCP TH wave is 6.12°±0.01° and is close to the theoretical value of 6.58°. According to the intensity profile of the measured TH diffraction pattern, the obvious zeroth-order diffraction is believed to be contributed by nonlinear polarization $3\varepsilon_0 \chi_{LLLR}^{(3)\text{eff}} E^R (E^L)^2$ (co-polarized) and $3\varepsilon_0 \chi_{RRRL}^{(3)\text{eff}} E^L (E^R)^2$ (cross-polarized). According to the simulation results shown in Figure 4(b), the cross-polarized total FW field confined inside the C4 silicon structure is weaker than the co-polarized total FW field, therefore, the intensity difference between the zeroth-order diffraction of co- and cross-polarized TH signals can be well understood. In addition, the asymmetric line shape of the TH intensity profile when switching the FW spin might be attributed to the low fabrication quality of the C4 sample (see the inset SEM image).

**2.5. Nonlinear holography based on silicon geometric-phase metasurfaces**

A promising application of the nonlinear metasurfaces is the holographic image reconstruction at a different wavelength than the FW (see Figure 7(a)). To demonstrate such potential by the silicon metasurface for the TH signal, we designed and encoded a phase-only hologram into the orientations of the nanofins. Figure 7(b) shows the SEM image of the fabricated sample, which is designed for generating a k-space hologram at the TH wavelength. The phase hologram is encoded with the geometric-phase proportional to two times of the rotational angle. Therefore, for the TH signals carrying the geometric-phase proportional to the four times of the rotational angle, no image can be obtained (see Figure 7(c)). By tuning the FW wavelength to 1300 nm and setting the average laser power at 300 mW, a clear nonlinear holographic image can be observed in our experiment. Figure 7(d) and 7(e) show the measured holographic image at the TH signals (433 nm wavelength) when the metasurface sample is illuminated by either an LCP or an RCP FW. Note that for RCP illumination the conjugated image is observed due to the sign change in the phase factors.

For the co-polarized TH signal, one can observe a clear holographic image while nothing can be observed for the cross-polarized TH signals. The results shown in Figure 7(d) can be understood with Equation 2. When the all-dielectric metasurface sample is illuminated with the LCP FW at 1300 nm, strong field confinement can be achieved in the nanofins. Therefore, the LCP TH signal mainly arises from the nonlinear polarizations $\varepsilon_0 \chi_{LLLL}^{(3)\text{eff}} e^{2i\theta} (E^L)^3$ and $\varepsilon_0 \chi_{LLLR}^{(3)\text{eff}} E^R (E^L)^2$. Hence, we can observe a bright target holographic image reconstructed by the TH signal carrying the geometric-phase of $\exp(2i\theta)$, a strong zeroth-order TH diffraction spot (carrying no geometric-phase information). Similarly, the RCP TH signal is mainly governed by $\varepsilon_0 \chi_{RLLL}^{(3)\text{eff}} e^{4i\theta} (E^L)^3$ and $\chi_{RRRL}^{(3)\text{eff}} E^L (E^R)^2$. Because the RCP TH signal that is carrying a geometric-phase of $\exp(4i\theta)$ is not designed to reconstruct the target image, only the terms that carrying a geometric-phase of $\exp(2i\theta)$ can generate the holographic image. Therefore, no image can be observed in the cross-polarized TH signals.

Based on the C1 and C4 structures, we further design and fabricate corresponding nonlinear hologram samples. For the hologram metasurface made of C1 silicon nanofins, Figure 8 (a) shows a theoretical calculation of the reconstructed field based on the co-polarized TH waves while the cross-polarized TH waves only present random noises readout, here the white circle refers to the boundary of the detectable k-space in the experiment. Figure 8(b) shows the measured holographic image when the metasurface sample is illuminated with an LCP FW whose wavelength and laser power are 1280 nm and 240 mW. A clear target image can be read from LCP TH signals. For the hologram metasurface composed by C4 silicon nanofin, the observable TH signals are only cross-polarized TH waves carrying the geometric-phase of $\exp(\pm 4i\theta)$. Figure 8(c) shows the theoretical calculation of the field reconstructed with the designed phase hologram whose phase value is encoded with four times of local rotational angle of silicon nanofin. We still set the laser power and wavelength of the FW as 240 mW and 1280 nm, Figure 8(d) shows the measured nonlinear hologram when metasurface sample constructed with C4 silicon nanofin is illuminated by LCP and RCP FW. Due to the low TH conversion efficiency and low diffraction efficiency of the C4 silicon we used in our work, the measured target hologram image is very weak and indistinct, however, we can still distinguish it from the random background given by co-polarized TH signals. For the LCP FW input and RCP TH output, we add the dot contour to help visualize the reconstructed target image. When we flip the spin of input FW, we can observe the conjugated image in the LCP TH signals as well (shown without the dot contour).

The geometric-phases associated with the TH waves we discuss here are based on a silicon nanofin array which involves the contribution of the coupling among the neighboring units. However, numerical simulations show that for an individual silicon nanofin with C1, C2 or C4 in-plane rotational symmetry without the coupling effects, the TH waves carry the spin-dependent geometric-phase modulation expected by the nonlinear geometric phase (Supporting Information). The influence of the coupling effects among the neighboring units on the deviation of the nonlinear phase can be understood as follows: firstly, the FW field confined inside the silicon nanofin will be reshaped when counting the coupling effects; secondly, the scattering of the side lobe of the TH field radiated from the dielectric nanofins will distort the TH far-field and forms the high order diffractions. In addition, our developed theory can also be generalized and expanded to the second-harmonic generation (SHG) and other higher-order harmonic generation processes.

## 3. Conclusion

In conclusion, we theoretically and experimentally study the nonlinear geometric-phases associated with the THG process occurring in all-dielectric silicon metasurfaces. The THG process involves the wave coupling among the different components of the FW field excited inside the dielectric nanoresonators and thereby gives rise to numerous nonlinear optical processes carrying different geometric-phases. We show that the dielectric nanofin structures with C1, C2, and C4 in-plane rotational symmetry generate circularly polarized TH signals in forward direction carrying the geometric-phases as predicted by the selection rule for nonlinear processes. Furthermore, we experimentally demonstrate nonlinear k-space holography by using all-dielectric geometric-phase metasurfaces composed by the silicon nanofins with C1, C2, and C4 in-plane rotational symmetry. The experiment results agree with our theoretical model. Our work offers a simple physical picture for the understanding of the nonlinear optical process occurring in all-dielectric geometric-phase metasurface and shows great application potential in nonlinear nanophotonics.

## 4. Experimental Section

*Full-wave simulation:* To figure out the transmitted phase and THG conversion efficiency of the silicon nanofin array, here we conducted the full-wave simulations by using the finite element method in the frequency domain. First, we calculated the total

FW field spatial distribution inside the nanofin. Next, the free space TH field was calculated based on the nonlinear polarization defined by the pre-calculated total FW field. In our simulation, the amplitude of the FW electric field was set as $10^8$ V/m, and the third-order nonlinear susceptibilities $\chi^{(3)}_{xxxx}$, $\chi^{(3)}_{yyyy}$, and $\chi^{(3)}_{zzzz}$ were assumed to be $2.45\times10^{-19}$ m$^2$ V$^{-2}$.[18] In addition, the third-order nonlinear susceptibilities $\chi^{(3)}_{iijj}$ $\chi^{(3)}_{ijij}$, and $\chi^{(3)}_{ijji}$ ($i \neq j$) were assumed to be one-third of $\chi^{(3)}_{xxxx}$, where $i$ and $j$ refer to the cartesian coordinates x, y, and z. The complex refractive index of the amorphous silicon (α-Si) we utilized in the simulation was adopted from the experimental measurement values of our laboratory fabricated α-Si film. And the refractive index of low-refractive-index silicon dioxide (SiO$_2$) was adopted from data measured by Maliston.[40] Periodic boundary condition was applied to calculate the far-field property of the infinite silicon nanofin array and the gradient geometric-phase metasurface.

*Hologram design:* The phase-only hologram is designed by using the Gerchberg-Saxton (GS) algorithm. As a kind of iterative phase retrieval algorithm, the GS algorithm does not calculate the wavefront in the hologram plane directly but constructs an iterative loop between the object plane and the hologram plane via a propagating function, such as the Fourier transform. During the process of iteration, the phase profile of the hologram is optimized with amplitude replacement applied on the object plane and the amplitude normalization applied on the hologram plane. In the nonlinear harmonic generation case, the phase information is calculated through such a method and is converted to the azimuthal angle distribution afterward, according to the nonlinear geometric-phase principle.

*Sample fabrication:* The all-dielectric silicon metasurfaces were fabricated on a glass substrate following the processes of deposition, patterning, lift-off, and etching. First, through plasma-enhanced chemical vapor deposition (PECVD), we prepared a 620-nm-thick amorphous silicon (α-Si) film. Following this, a poly-methyl methacrylate resist layer was spin-coated onto the a-Si film and baked on a hot plate at 170 °C for 2 min to remove the solvent. Next, the desired structures were patterned by using standard electron beam lithography. Subsequently, the sample was developed in 1:3 MIBK:IPA solution and then washed with IPA before being coated with a 20-nm-thick chromium layer by electron beam evaporation. Afterward, a lift-off process in hot acetone was performed. Finally, by using inductively coupled plasma reactive ion etching (ICP-RIE), the desired structures were transferred from chromium to silicon.

*Optical Characterization:* The nonlinear response of the metasurface is measured by an optical setup. The metasurface is illuminated by a circularly polarized femtosecond laser beam between 1200 nm and 1400 nm. The laser source is a Ti:sapphire femtosecond laser pumped optical parametric oscillator (OPO) with a typical pulse length of 200 fs, a repetition rate of 80 MHz and a typical output power of 300 mW. The incident laser beam was focused onto the metasurface (f = 50 mm, beam width: ~50 µm) in order to enhance the nonlinear response of the metasurface. The generated third harmonic light from the metasurface is collected by a microscope objective (40x magnification, NA = 0.6), whose back focal plane is imaged onto an sCMOS camera (Andor Zyla 4.2). To image the back focal plane onto the camera, we used two lenses with focal distances of $f = 150\,\text{mm}$. To distinguish between different circular polarization states, we used a combination of a quarter-wave plate and a linear polarizer. Optical filters are used to suppress the fundamental light. The diffraction spots and the holograms reconstruct at the back focal plane of the microscope objective, which allows for a unified measurement setup for the gradient and hologram metasurfaces.

**Acknowledgments**

Special thanks to Ms. Yuhong Na (L) for rendering high-quality artistic figures. This project has received funding from the European Research Council (ERC) under the European Union's Horizon 2020 research and innovation programme (grant agreement No 724306) and the Deutsche Forschungsgemeinschaft (Grant No. DFG TRR142/C05). L. H acknowledges the support from the Beijing Outstanding Young Scientist Program (BJJWZYJH01201910007022) and the National Natural Science Foundation of China program (No. 61775019).

**References**

[1]　N. Yu, F. Capasso, *Nat. Mater.* **2014**, *13*, 139.
[2]　A. V. Kildishev, A. V. Boltasseva, V. M. Shalaev, *Science* **2013**, *339*, 123009.
[3]　N. Yu, P. Genevet, M. A. Kats, F. Aieta, J.-P. Tetienne, F. Capasso, Z. Gaburro, *Science* **2011**, *334*, 333.
[4]　X. Ni, S. Ishii, A. V. Kildishev, V. M. Shalaev, *Light: Sci. Appl.* **2013**, *2*, e27.
[5]　M. Khorasaninejad, W. Chen, R. Devlin, J. Oh, A. Y. Zhu, F. Capasso, *Science* **2016**, *352*, 1190.
[6]　L. Huang, X. Chen, H. Mühlenbernd, H. Zhang, S. Mei, B. Bai, Q. Tan, G. Jin, K. Cheah, C. Qiu, J. Li, T. Zentgraf, S. Zhang, *Nat. Commun.* **2013**, *4*:2808.
[7]　G. Zheng, H. Mühlenbernd, M. Kenney, G. Li, T. Zentgraf, S. Zhang, *Nat. Nanotech.* **2015**, *10*, 308.
[8]　R. C. Devlin, A. Ambrosio, N. A. Rubin, J. P. B. Muller, F. Capasso, *Science* **2017**, *358*, 896.
[9]　X. Chen, L. Huang, H. Mühlenbernd, G. Li, B. Bai, Q. Tan, G. Jin, C. Qiu, S. Zhang, T. Zentgraf. *Nat. Commun*. **2012**, *3*:1198.
[10]　X. Yin, Z. Ye, J. Rho, Y. Wang, X. Zhang, *Science* **2013**, *339*, 1405.
[11]　L. Huang, X. Chen, B. Bai, Q, Tan, G. Jin, T. Zentgraf, S. Zhang, *Light: Sci.*

*Appl.* **2013**, *2*, e70.

[12]     J. P. B. Muller, N. A. Rubin, R. C. Devlin, B. Groever, F. Capasso, *Phys. Rev. Lett.* **2017**, *118*, 113901.

[13]     C. Pfeiffer, A. Grbic, *Phys. Rev. Lett.* **2013**, *110*, 197401.

[14]     M. Kauranen, A. V. Zayats, *Nat. Photon.* **2012**, *6*, 737.

[15]     G. Li, S. Zhang, T. Zentgraf, *Nat. Rev. Mater.* **2017**, *2*, 17010.

[16]     S. Kruk, Y. Kivshar, *ACS Photon.* **2017**, *4*, 2638.

[17]     A. I. Kuznetsov, A. E. Miroshnichenko, M. L. Brongersma, Y. S. Kivshar, B. Luk'yanchuk, *Science* **2016**, *354*, 6314.

[18]     W. Zhao, H. Jiang, B. Liu, J. Song, Y. Jiang, C. Tang, J. Li, *Sci. Rep.* **2016**, *6*, 30613.

[19]     D. Smirnova, Y. S. Kivshar, *Optica* **2016**, *3*, 1241.

[20]     D. Rocco, V. F. Gili, L. Ghirardini, L. Carletti, I. Favero, A. Locatelli, G. Marino, D. N. Neshev, M. Celebrano, M. Finazzi, G. Leo, C. D. Angelis. *Photonics Res.* **2018**, *6*, B6.

[21]     L. Wang, S. Kruk, K. Koshelev, I. Kravchenko, B. Luther-Davies, Y. Kivshar, *Nano Lett.* **2018**, *18*, 3978.

[22]     Y. Yang, W. Wang, A. Boulesbaa, I. I. Kravchenko, D. P. Briggs, A. Puretzky, D. Geohegan, J. Valentine, *Nano Lett.* **2015**, *15*, 7388.

[23]     A. S. Shorokhov, E. V. Melik-Gaykazyan, D. A. Smirnova, B. Hopkins, K. E. Chong, D. Choi, M. R. Shcherbakov, A. E. Miroshnichenko, D. N. Neshev, A. A. Fedyanin, Y. S. Kivshar, *Nano Lett.* **2016**, *16*, 4857.

[24]     S. Campione, S. Liu, L. I. Basilio, L. K. Warne, W. L. Langston, T. S. Luk, J. R. Wendt, J. L. Reno, G. A. Keeler, I. Brener, M. B. Sinclair, *ACS Photon.* **2016**, *3*, 2362.

[25]     P. P. Vabishchevich, S. Liu, M. B. Sinclair, G. A. Keeler, G. M. Peake, I. Brener, *ACS Photon.* **2018**, *5*, 1685.

[26]     G. Grinblat, Y. Li, M. P. Nielsen, R. F. Oulton, S. A. Maier. *ACS Nano* **2017**, *11*, 953.

[27]     Y. Yang, V. A. Zenin, S. I. Bozhevolnyi, *ACS Photon.* **2018**, *5*, 1960.

[28]     L. Xu, M. Rahmani, K. Z. Kamali, A. Lamprianidis, L. Ghirardini, J. Sautter, R. Camacho-Morales, H. Chen, M. Parry, I. Staude, G. Zhang, D. Neshev, A. E. Miroshnichenkov. *Light: Sci. Appl.* **2018**, *7*:44.

[29]     E. Almeida, O. Bitton, Y. Prior, *Nat. Commun.* **2016**, *7*:12533.

[30]     Y. Gao, Y. Fan, Y. Wang, W. Yang, Q. Song, S. Xiao, *Nano Lett.* **2018**, *18*, 8054.

[31]     J. Lee, M. Tymchenko, C. Argyropoulos, P. Chen, F. Lu, F. Demmerle, G. Boehm, M. Amann, A. Alù, M. A. Belkin, *Nature* **2014**, *511*, 65.

[32]     M. Tymchenko, J. S. Gomez-Diaz, J. Lee, N. Nookala, M. A. Belkin, A. Alù,

*Phys. Rev. Lett.* **2015**, *115*, 207403.

[33] M. Tymchenko, J. S. Gomez-Diaz, J. Lee, N. Nookala, M. A. Belkin, A. Alù, *Phys. Rev. B* **2016**, *94*, 214303.

[34] G. Li, S. Chen, N. Pholchai, B. Reineke, P. Wong, E. Pun, K. Cheah, T. Zentgraf, S. Zhang, *Nat. Mater.* **2015**, *14*, 607.

[35] S. Chen, G. Li, F. Zeuner, W. Wong, E. Pun, T. Zentgraf, K. Cheah, S. Zhang, *Phys. Rev. Lett.* **2014**, *113*, 033901.

[36] K. Konishi, T. Higuchi, J. Li, J. Larsson, S. Ishii, M. Kuwata-Gonokami, *Phys. Rev. Lett.* **2014**, *112*, 135502.

[37] Y. Tang, Y. Intaravanne, J. Deng, K. Li, X. Chen, G. Li, *Phys. Rev. Appl.* **2019**, *12*, 024028.

[38] W. Ye, F. Zeuner, X. Li, B. Reineke, S. He, C. Qiu, J. Liu, Y. Wang, S. Zhang, T. Zentgraf, *Nat. Commun.* **2016**, *7*:11930.

[39] B. Reineke, B. Sain, R. Zhao, L. Carletti, B. Liu, L. Huang, C. Angelis, T. Zentgraf, *Nano Lett.* **2019**, *19*, 6585.

[40] I. H. Malitson. *J. Opt. Soc. Am.* **1965**, *55*, 11205.

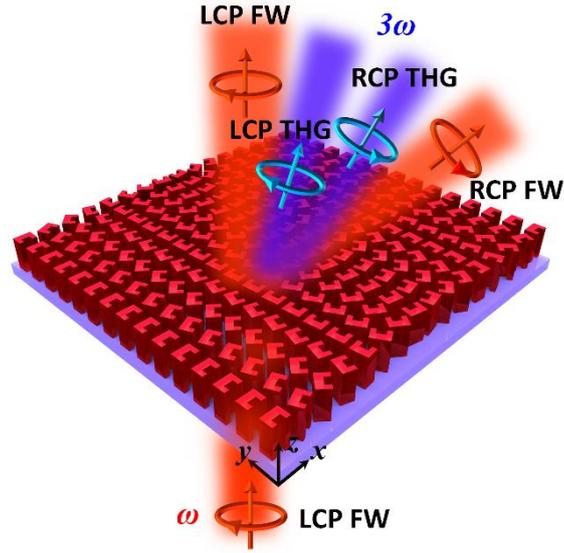

**Figure 1.** Schematic of the anomalous nonlinear beam steering of the all-dielectric gradient nonlinear geometric-phase metasurface. The silicon nanofins are placed on the silica substrate and the surrounding medium is selected as air. The LCP FW at $\omega$ illuminates the metasurface from the substrate and generates the RCP and LCP TH beams at $3\omega$ with different anomalous refraction angles.

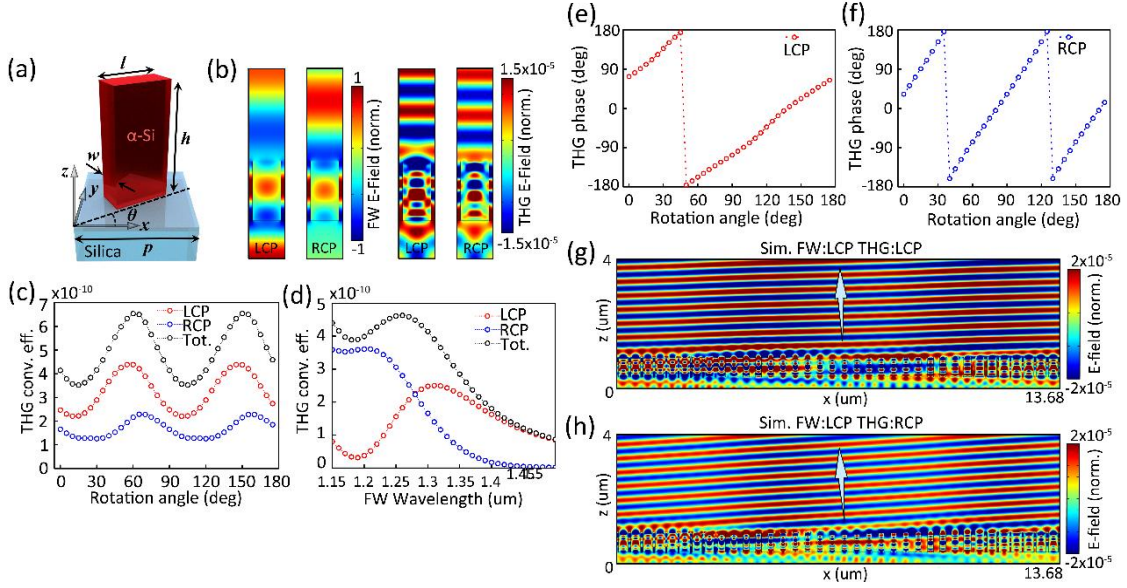

**Figure 2.** Nonlinear response of the locally rotated silicon nanofin array. (a) Schematic of the silicon nanofin with C2 in-plane rotational symmetry and its geometric parameters. (b) The LCP and RCP components of the FW field at 1300 nm and the corresponding TH field of the silicon nanofin array, here the rotated angle is selected as 0°. All fields are normalized with input FW electric field amplitude. (c) The LCP (red), RCP (blue) and total (black) forward THG conversion efficiency of the rotated silicon nanofin array ranging from 0° to 180° with an angular rotating step of 5°. (d) The LCP (red), RCP (blue) and total (black) forward THG conversion efficiency of the rotated silicon nanofin array by sweeping the FW wavelength from 1150 nm to 1500 nm. Numerical calculation of the spin-dependent anomalous refractions of the TH signals generated from the all-dielectric nonlinear gradient geometric-phase metasurface. Selecting the LCP FW at 1300 nm, the simulated nonlinear geometric-phase of the (e) LCP and (f) RCP TH signal when rotating the silicon nanofin from 0° to 180° with an angular rotation step of 5°. The simulated free space field distribution of the (g) LCP and (h) RCP TH signal. All fields are normalized with input FW electric field intensity.

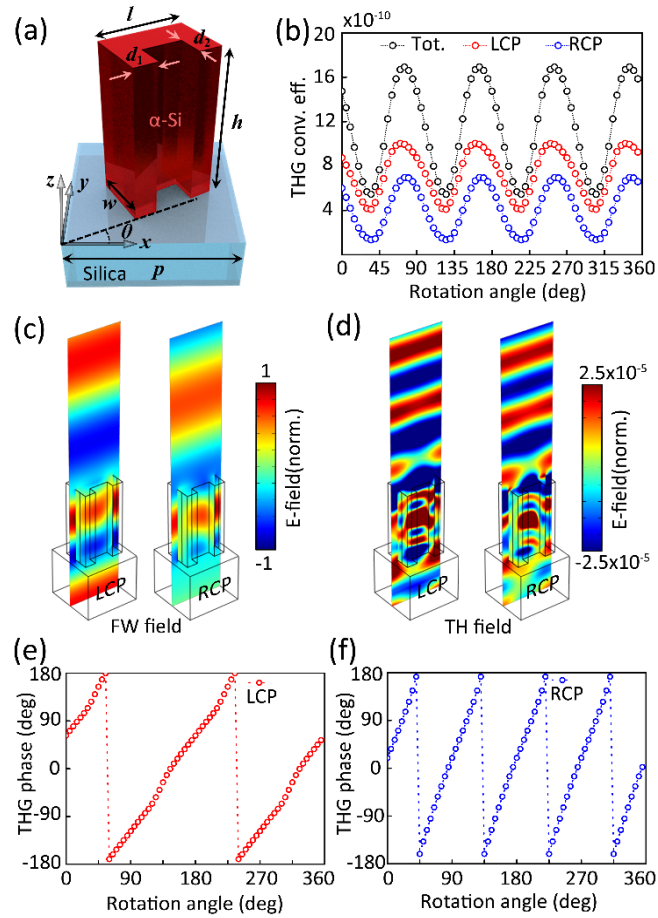

**Figure 3.** Numerical calculation of the TH signals generated from the all-dielectric geometric-phase metasurface composed by locally rotated silicon nanofin with C1 rotational symmetry. (a) Schematic of the silicon nanofin and its geometry parameters. Selecting the LCP FW at 1240 nm, the simulated (b) THG conversion efficiency when gradually rotating the silicon nanofin from 0° to 360° with an angular rotation step of 5°. Simulated field distribution of LCP and RCP components of the (c)total FW fields and (d)TH fields. All fields are normalized with input FW electric field intensity. Nonlinear geometric-phase of the (c) LCP and (d) RCP TH wave by rotating the silicon nanofin from 0° to 360° with an angular rotation step of 5°.

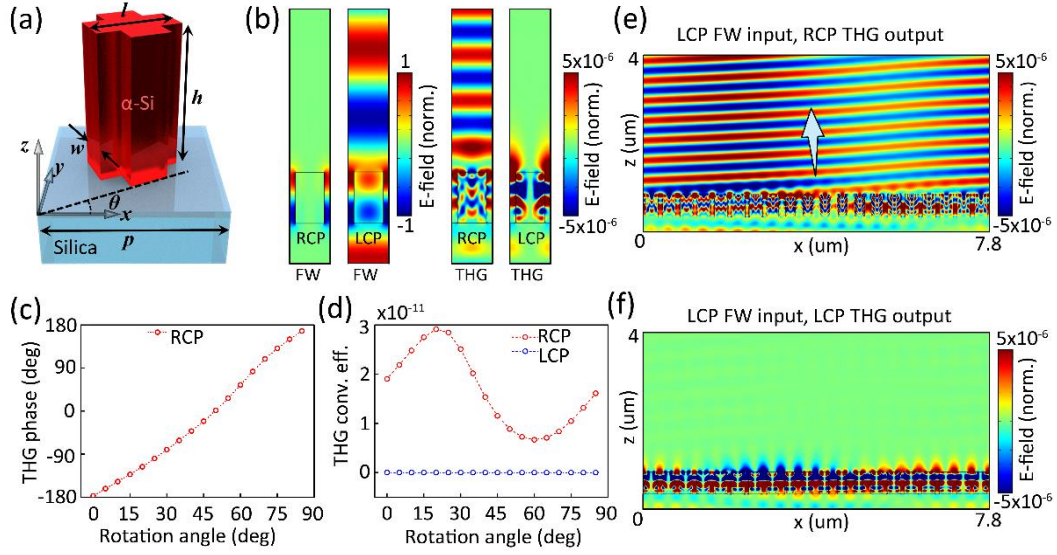

**Figure 4.** Numerical calculation of the nonlinear beam steering of the TH waves generated from the dielectric geometric-phase metasurface composed by silicon nanofin of C4 in-plane rotational symmetry. (a) Schematic of the silicon nanofin. Selecting the LCP FW wavelength as 1350 nm, the field distribution of (b) different circularly polarized components of the FW and TH field excited in the silicon nanofin array whose local rotational angle $\theta$ is 0°. The nonlinear geometric-phase of (c) the RCP TH signal and (d) the THG conversion efficiency of the RCP (red circle) and LCP (blue circle) TH waves when rotating the silicon nanofins from 0° to 90° with an angular rotation step of 5°. The simulated free space TH field distribution of (e) the anomalous refracted RCP TH waves and (f) surface bounded LCP TH waves. All fields are normalized with input FW electric field intensity.

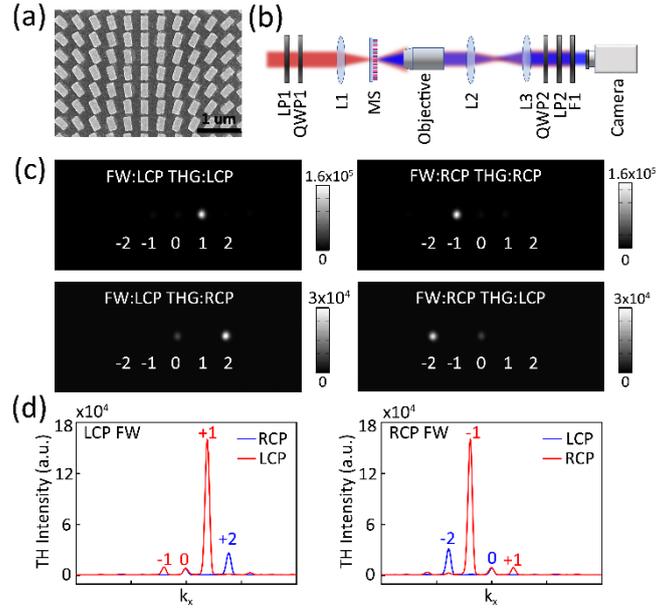

**Figure 5.** Experimental verification of the TH signals generated from the all-dielectric geometric-phase metasurface with a spin-dependent surface phase gradient. (a) SEM image of the all-dielectric gradient geometric-phase metasurface sample, the silicon nanofin with C2 in-plane symmetry is rotated by the step of 9°. The scale bar is 1 um. (b) Schematic of the experimental setup. F1 refers to the short-pass filter; LP1 and LP2 refer to linear polarizer; QWP1 and QWP2 refer to the quarter waveplate; L1, L2, and L3 refer to the focusing lens. (c) The measured TH diffraction patterns under the illumination of the circularly polarized FW at 1200 nm. (d) Intensity profile of center cutline of the measured TH diffraction pattern.

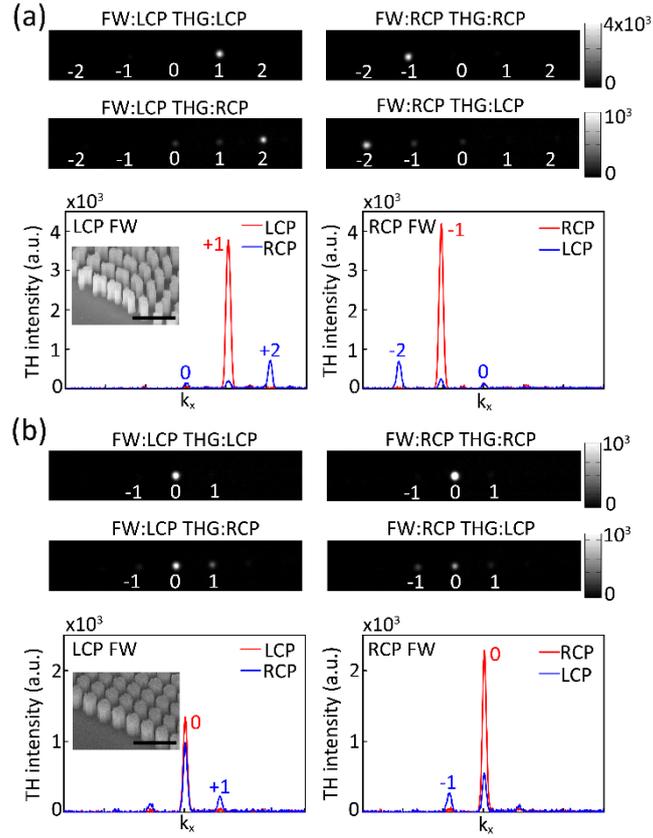

**Figure 6.** Measured diffraction pattern and center cutline intensity profile of TH waves radiated from the gradient geometric-phase metasurface composed by (a) C1 and (b) C4 silicon nanofins. Scale bar of inset SEM images of gradient geometric-phase metasurface is 1 um.

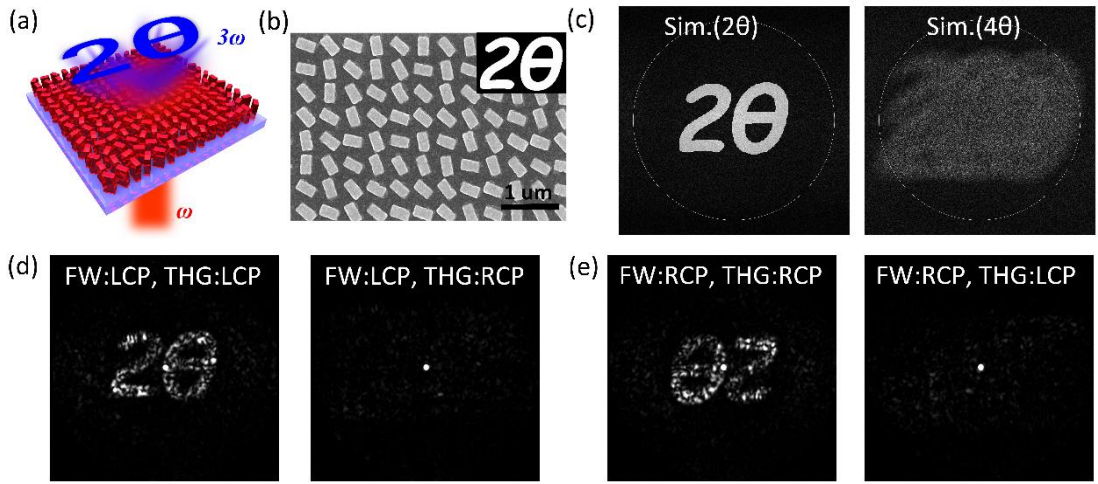

**Figure 7.** Nonlinear holography based on the nonlinear all-dielectric geometric-phase metasurface. (a) Artistic sketch of the nonlinear holography realized by the all-dielectric geometric-phase metasurface. (b) SEM image of the metasurface sample and the corresponding target image (inset). (c) Theoretical calculation of the reconstructed field based on the readout signals carrying the geometric-phase of $\exp(2i\theta)$ (left) and $\exp(4i\theta)$ (right), the white circle refers to the boundary of k-space. Measured nonlinear hologram when the metasurface is illuminated with the (d) LCP and (e) RCP FW.

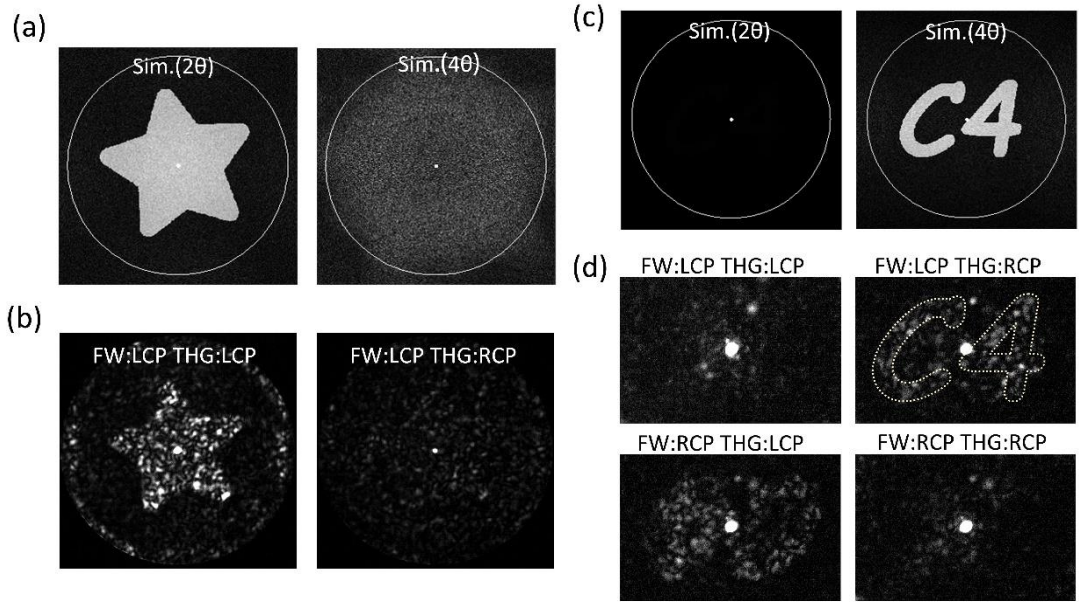

**Figure 8.** Nonlinear holography based on C1 and C4 silicon nanofins. For the sample made of C1 silicon nanofins, (a) theoretical calculation of the reconstructed field based on the readout signals carrying the geometric-phase of $\exp(2i\theta)$ (left) and $\exp(4i\theta)$ (right), and the (b) measured nonlinear hologram reconstructed from LCP and RCP TH signals when the sample is normally illuminated with an LCP FW. For the sample made of C4 silicon nanofins, (c) theoretical calculation of the reconstructed field based on the readout signals carrying the geometric-phase of $\exp(2i\theta)$ (left) and $\exp(4i\theta)$ (right), and the (d) measured nonlinear hologram when the metasurface sample is normally illuminated with an LCP and RCP FW, respectively.